# A Provenance Framework for Policy Analytics in Smart Cities

Barkha Javed, Richard McClatchey, Zaheer Khan, and Jetendr Shamdasani

*Faculty of Environment and Technology, University of the West of England, Frenchay Campus, Bristol, UK*
{barkha.javed, richard.mcclatchey, Zaheer2.Khan}@uwe.ac.uk, Jetendr.Shamdasani@cern.ch

Abstract: Sustainable urban environments require appropriate policy management. However, such policies are established as a result of underlying, potentially complex and long-term policy making processes. Consequently, better policies require improved and verifiable planning processes. In order to assess and evaluate the planning process, transparency of the system is pivotal which can be achieved by tracking the provenance of policy making process. However, at present no system is available that can track the complete cycle of urban planning and decision making. We propose to capture the complete process of policy making and to investigate the role of Internet of Things (IoT) provenance to support design-making for policy analytics and implementation. The environment in which this research will be demonstrated is that of Smart Cities whose requirements will drive the research process.

## 1 INTRODUCTION

Unprecedented rapid urbanisation has been observed in recent years; according to the World Health Organization (2016), today's urban population accounts for more than 50% of the total global population and is expected to further increase to 70% by 2050 (British Standard Institute, 2014). The consequent growing population is placing pressure on social, environmental and other resources including the wider city infrastructure. To meet these challenges, the notion of Smart Cities has emerged in recent years. Smart Cities are often referred to as the use of Information and Communication Technology (ICT) to improve the quality of life and provide sustainable living for citizens (Bakici et al., 2013). However, the vision of Smart Cities is only possible with new and better approaches to urban planning and decision making (Chourabi et al., 2012; British Standard Institute, 2014). Nevertheless, urban planning and decision making is a challenging task as it entails diverse information, complex processes, and involves various stakeholders.

The policy making process consists of different stages including problem identification, agenda setting, analysis, negotiation and decision making, implementation, and evaluation (Khan, 2014). Each stage has further associated tasks; for example, the problem identification stage may encompass the acquisition of quantitative and qualitative information, potentially through the use of the Internet of Things (IoT), city databases, etc. The data analysis may involve the investigation of data and evidence, the assessment of alternative scenarios, and the identification of the cause of the issue(s): for example, the identification of the cause of air pollution in a city using information gathered from IoT sensors. Similarly, other planning phases also consist of further tasks. In addition, various stakeholders are involved at each phase of the process. New emerging trends in policy require processes to adopt more transformational approaches (e.g. bottom-up initiatives) to enable collaborative decision making. A generic planning cycle should support both top-down and bottom-up planning initiatives. In recent years, use of new ICT solutions for public participation (i.e. participatory sensing) has transformed planning processes in smart cities (Batty et al., 2012).

The effectiveness of urban policies is largely dependent on evidence employed and decisions taken during the process. Tracking of the complete lifecycle of policy making is required in the planning process for evaluation of decisions and the evidence used in policy making. This helps to achieve more informed policy decisions and to make the system more transparent, legitimate and accountable (Coglianese et al., 2008; Jeannine and Sabharwal, 2009). Recent literature indicates that local governments are realising the potential of co-creativity and co-production through multi-stakeholder participation in planning processes and hence are looking for open governance models where transparency in these processes plays an important role.

In order to track the processes and decision-making in the policy cycle, extensive provenance information should be collected (Ram and Liu, 2009). This provenance includes information about how, when, by whom and why data has been gathered, analysed and used in policy making. Such information is essential in order to track the complete policy cycle and to provide data integration and accountability at each decision making stage. Provenance for smart cities can provide useful information such as how and when data was collected, the ownership of that data, how collected information is being processed, evidence considered during planning, potential stakeholders in the planning, how citizens' feedback were accommodated, decisions made during planning, alternatives considered during decision making, and the outcome of decisions (Lopez-de-Opina et al., 2013). Therefore the first motivation for this research is to employ provenance information to track all the phases of the smart city policy cycle where each phase may be considered as an individual system having associated inputs, outputs, sets of tasks, and different stakeholders' involvement. Tracking of each phase will provide substantial information for subsequent phases which will guide further planning process.

The provenance tracked during policy making process can be used for further analysis. For example, provenance can provide a reference for future planning. Collected evidence can be exploited to explore the success or failure factors of previously devised policy which can provide guidance for devising other/future policies. It can also be used to analyse the impact of policy decisions on other associated city operations. The analysis of provenance can thus improve future decisions. Therefore, the second motivation of this study is to investigate the role of provenance to support policy analytics in smart cities environment.

Urban planning entails diverse and city-wide information (such as transportation, air quality monitoring, health, waste management etc). A flexible system is required to capture the diverse information associated with urban planning and to track the complete history of changes. In this regard, model-driven engineering (MDE) can be applied to develop a provenance framework to support policy analytics. The provenance framework will also investigate the description driven approach such as that of the CRISTAL ( McClatchey et al., 2014) to capture provenance in the smart cities domain. This research will evaluate the suitability of model driven approaches to support provenance information gathering for smart cities by experimentation. Furthermore, we are aware of the issues related to security of provenance data. However, this is out of scope of this research.

This paper presents the needs and benefits for tracking the full lifecycle of the planning process for smart cities. The overall aim of this research is to investigate the extent to which it is possible to make effective use of provenance for evidence based policy analytics. The next section outlines related work and a provenance framework for policy analytics is presented thereafter before conclusions and future work are subsequently outlined.

## 2 LITERATURE REVIEW

### 2.1 A Planning Model for Smart Cities

The functioning of a city is a reflection of its underlying planning and decision making. Traditional approaches to urban planning can benefit from new ICT solutions to address the growing challenges of recent rapid urbanization. To address the issues related to traditional approaches, smart cities modelling considers using new operating models for the planning of cities (Horelli and Wallin, 2013; British Standard Institute, 2014). Smart cities planning encourage an open and transparent governance process and facilitate public participation in a planning process.

Open Government (Geiger and Lucke, 2011) is a mechanism employed in recent years for government accountability and public scrutiny. This approach provides transparency in the governance system by enabling the availability of government-held information to the public; furthermore, this system also encourages citizens' participation. Different projects to support the Open Government approach have been initiated in the past few years. Of which Urban API (2014), FUPOL (2016) and Smarticipate ( 2016) are suitable examples.

Open government provides transparency but may not necessarily ensure reliability and trust in a system (Ceolin et al., 2013). This can be achieved by employing provenance for tracking the planning process. The suitability of provenance for tracking urban planning and decision making is further discussed in the subsequent section.

## 2.2 Provenance for Tracking Planning Processes in Smart Cities

. Provenance is often employed to trace the audit trail and usage of data, to estimate data quality and reliability and accuracy, to verify the validity of information, integrity, authenticity, replication and repetition of data and processes, to validate the attribution of data, and to establish transparency and trust in the system (Carata et al., 2014; Simmhan et al., 2005).

The significant dual challenges of gathering and storing provenance data in complex Smart Cities has motivated a number of research efforts in recent years. d'Aquin et.al (2014) addresses the management of diverse datasets produced by different objects in Smart Cities. Provenance is employed in (Lopez-de-Opina et al., 2013; Emaldi et al. 2013) for addressing validation and trust issues related to open data in Smart Cities. Provenance is employed in (Packer et al., 2014) for transparency and accountability of sharing services in smart cities.

The literature demonstrates the potential use of provenance in Smart City environments. However, utilising provenance information and data emerging from IoT sensor nets to capture the processes needed in the planning process in smart cities has not yet been investigated. Nevertheless, the suitability of provenance for urban planning has been discussed by Edwards et al (2009). Furthermore, eSocialScience tools and techniques have been proposed to support social scientists involved in policy-related research. Evidence-based policy simulation is a focus of the OCOPOMO project (Lotzmann and Wimmer, 2012) which enables policy formulation using a set of ICT tools. The tools facilitate policy makers in modelling policies and in communicating them to other stakeholders for feedback. Scherer (2015) extends the OCOPOMO project by using a model-driven approach in the project.

What is required is a holistic approach to managing the full lifecycle of policy making for smart cities. This will necessitate the use of a process oriented approach to identify socio-technical activities and exchange of data among actors in a policy cycle. This approach will deal with the integration of heterogeneous data in a common conceptual model (potentially description-driven, as in the CRISTAL software) and the gathering, curation and analysis of data emerging from smart city sensing devices plus tracking the provenance and processing of those data and how they may influence decision making, policy implementation and its evaluation in a city-wide environment.

The existing work (Edwards et al., 2009; Lotzmann and Wimmer, 2012; Scherer et al., 2015) shows the potential role of provenance in urban planning. However, the current systems do not track all activities of the policy cycle and are not in the context of smart cities. Citizens' participation is important for smart cities planning ( BristolisOpen, 2015). Therefore, provenance gathering will also need to capture how their suggestions were accommodated in the policy process. Provenance tracking of smart cities' planning will provide a rich source of information regarding the policy making process. This information can be used to support policy analytics in smart cities which is discussed in section 2.3.

## 2.3 Using Provenance to Support Policy Analytics in Smart Cities

Policy analytics in the past couple of years has attracted the attention of many researchers (De Marchi et al., 2012; Tsoukiàs et al., 2013; Daniell et al., 2015). Opinion mining has been employed by Kaschesky et al (2011) to track and analyse the citizens' participation in policy making process. Similarly, possible use of preference learning, text mining, value-driven analysis, prospective analysis, and data mining for policy analytics has been specified by a number of researchers (Tsoukiàs et al., 2013; Daniell et al., 2015).

The planning process requires both data and value-driven decision making (Tsoukiàs et al., 2013). Therefore, in order to enable policy analytics and to aid in decision making, tracking of both data and values is required. Provenance of the policy making process will provide an integrated platform and will provide rich information regarding the process such as the evidence used (in the case of smart cities, data from IoT sensors is pertinent), public engagement, and decisions of policy makers. Such information can be used for analysis and to inform current and future decision making.

Provenance can be employed to find useful information and can be used for the purpose of learning and knowledge discovery (Liu et al., 2013). Huynh et al (2013) used provenance analytics to assess the quality of the crowd-generated data. Margo and Smogor (2010) employed machine learning classification techniques in order to classify files using their provenance. Huynh and Margo show the possible use of provenance to support data analytics. Provenance can also be considered to support policy analytics in the smart cities planning process. However, the potential benefits and use of provenance for policy analytics need to be explored by further study and experimentation.

## 3 PROVENANCE FRAMEWORK TO SUPPORT POLICY ANALYTICS

Figure 1 (Khan, 2014) depicts a typical policy making process and aim is to capture provenance through all stages of this process.

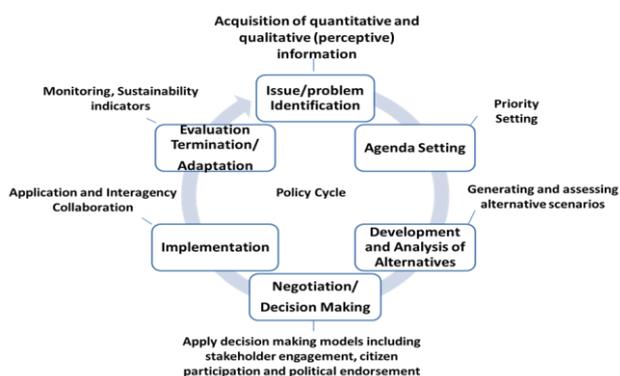

Figure 1: Policy Cycle (Khan, 2014)

Each phase of the planning process has further associated processes and tasks. This suggests that each stage can be considered as an individual system and points to a system-of-systems (Luzeaux and Ruault, 2010) approach to modelling the policy cycle. This will be investigated in the current research.

Our research will assess the extent to which provenance of planning processes are effective for policy making and analytics in smart cities and the framework needed to support these analytics. In order to clarify the provenance support for policy analytics, table 1 shows the provenance captured at each phase of the smart cities policy cycle and possible analytics that can be applied using provenance. However, this will be improved on further research.

### 3.1 Case Study: Air Quality Monitoring

For further clarification, let us consider the scenario of Air Quality Monitoring. Assume a large concentration of Carbon Monoxide (CO) and Nitrogen Oxide (NO) has been recorded in 'City A' which is mostly contributed by transportation fumes. In order to minimise the concentration of identified pollutants, a new policy is required to be put in place. To devise a new policy, air quality data is captured from air monitoring IoT sensors by analysts who run statistical analysis according to the thresholds set by the current air quality policy (for example EU and national levels). Similarly, in order to investigate the potential role of traffic in air pollution, information regarding traffic and vehicles has been collected from highway and vehicle licensing agencies respectively. The collected data is then analysed by the analyst and is communicated to the concerned department. Provenance information is gathered at the point of recording air quality, traffic, and vehicle data; its analysis and its outputs are recorded in order to facilitate linkage to the subsequent decision making stage.

Table 1: Policy analytics using provenance

| Policy cycle phases | Associated Tasks | Provenance Information | Possible Analytics Approaches |
|---|---|---|---|
| Survey and Problem identification | Acquisition of qualitative and quantitative data (data from sensors, surveys, interviewees etc) | Verifiable source(s) of data (social network, IoT, city databases, etc.), perception/views of different stakeholders | Range of analytic technique (such as data mining, machine learning algorithms), value-driven analysis |
| Agenda setting | Priority Setting | Domain experts' views | Value-based analysis |
| Analysis | Investigation of evidence, assessment of alternatives, identification of the cause of issues | Capture analysis techniques, capture evidence details, stakeholders' values, views/perceptions. | Different data analytics techniques along with value-based analysis, perception analysis |
| Decision making | Negotiation among stakeholders, citizens involvement, decision making based on policies/evidence | Stakeholders' perception, citizens opinion, capture evidences used in decision making, policy success indicators | Opinion analysis, conflict resolution, social learning capabilities |
| Implementation | Interagency cooperation (some metrics to track policy implementation) | Track the data used to assess implementation compliance to original policy specification | - |
| Evaluation | Monitor the policy | Track the matrices used for monitoring | Perception analysis, data analytics techniques |

For our case study let us now assume the analysis demonstrates the role of traffic in air pollution. The issue is communicated to urban planners, city administration, the environment agency, and citizens via the recorded provenance information in order to ensure the trustworthiness of the analysis. Based on the feedback of stakeholders, policy makers then propose a strategy to minimise traffic congestion by devising alternative routes. These decisions are also recorded alongside the processed data in the provenance store. If traffic exceeds a particular threshold then it is routed to other available routes. The proposed threshold is negotiated among policy makers. The strategy is implemented and air quality is continuously monitored to evaluate the policy based on some evaluation criteria. Each stage in the process is recorded in order to provide full traceability of the policy cycle.

This case study demonstrates the complex process of planning; the various evidence (data from IoT sensors, traffic and vehicles data, data gathered at each phase), decision choices (of policy makers and stakeholders), evaluation criteria (set by air quality policy in given example), and stakeholders (urban planners, city administration, the environment agency, and citizens) involvement in the process. For transparency in the system, provenance information is captured at each phase of policy of the planning process. Let us suppose that the devised policy is not successful. In order to uncover the issue, provenance at each stage can be carefully analysed (by using analytics techniques). The identified issue is addressed by considering options and therefore devising a new policy, driven by the model-based holistic policy support framework. Similarly, provenance can also provide assistance in evaluating accountability, exploring the benefits of public participation, evaluating decisions.

# 4 CONCLUSION AND FUTURE WORK

Urban planning and decision making is a challenging task as it entails complex processes, it involves various actors, and uses data collected from heterogeneous sources. To improve services and to guide in future decision making, all planning decisions are required to be maintained which necessitates the capturing of provenance for smart cities environments. This paper puts forward the idea of tracking the full urban planning process and considers each phase of the process as an individual system. Furthermore, the idea of using provenance, potentially using a description-driven, model-based approach to supporting policy analytics is also presented in this position paper.

Future work will consider the implementation and evaluation of the proposed research study using practical examples of data derived from IoT sensor network. The aim is to explore what possible analyses could be carried with Smart Cities provenance data. Policy analytics is an area which is still in its infancy as highlighted by the literature in section 2 of this paper; therefore this study will explore how policy analytics can be supported by using provenance captured during urban planning.

## REFERENCES

Bakici, T et al., (2013). A Smart City Initiative: The Case of Barcelona. *Journal of the Knowledge Economy. pp.*135-148.

Batty, M et al., (2012). Smart cities of the future. *European Physical Journal Special Topics*, pp. 481–518.

Bristol Is Open project. Available at: www.bristolisopen.com [Accessed 4 March, 2016].

British Standard Institute, (2014). PAS 181: 2014 Smart city framework – Guide to establishing strategies for smart cities and communities. United Kingdom. Available at: http://www.bsigroup.com/en-GB/smart-cities/Smart- Cities-Standards-and-Publication/PAS-181-smart-cities-framework/

Carata, L et al., (2014). A Primer on Provenance. *Commun. ACM*.

Ceolin, D et al., (2013). Reliability Analyses of Open Government Data. *In URSW*, pp. 34–39.

Chourabi, H et al., (2012). Understanding Smart Cities: An Integrative Framework. The *45th Hawaii International Conference on System Sciences*, pp. 2289 - 2297.

Coglianese, C et al., (2008). Transparency and Public Participation in the Rulemaking Process. *A *Nonpartisan Presidential Transition Task Force Report*.

Daniell, KA et al., (2015). Policy analysis and policy analytics. *Annals of Operations Research*., 10.1007/s10479-015-1902-9

d'Aquin, M et al., (2014). Dealing with Diversity in a Smart-City Datahub. In *Fifth Workshop on Semantics for Smarter Cities*. pp. 68–82.

De Marchi, G et al., (2012). From Evidence Based Policy Making to Policy Analytics. *Cahier du LAMSADE 319. Université Paris Dauphine, Paris*.

Edwards, P et al., (2009). esocial science and evidence-based policy assessment: challenges and solutions. *Social Science Computer Review*. vol. 27(4), pp. 553–568.

Emaldi et al., (2013). To trust, or not to trust: Highlighting the need for data provenance in mobile apps for smart cities. *International Workshop on Semantic Sensor Networks (SSN)*.

FUPOL. Available at: http://www.fupol.eu/en [Accessed 4 March, 2016]

Geiger, C. P and Lucke, J.V., (2011). Open Government Data. In CeDEM11. *Conference for E-Democracy and Open Government*. pp. 183–194.

Horelli,L and Wallin, S., (2013). New Approaches to Urban Planning Insights from Participatory Communities. *Aalto University Publication series* Aalto-ST 10/2013, pp.11-16.

Huynh, T.D., et al., (2013). Interpretation of crowdsourced activities using provenance network analysis. In:*First AAAI Conference on Human Computation and Crowdsourcing*.

Jeannine, E. R and Sabharwal, M., (2009). Perceptions of Transparency of Government Policymaking: A Cross-National Study. *Government Information Quarterly* 26(1): pp.148-157.

Kaschesky, M et al., (2011). Opinion mining in social media: modeling, simulating, and visualizing political opinion formation in the web. *In: International Conference on Digital Government Research*.

Khan, Z et al., (2014). ICT enabled participatory urban planning and policy development: The UrbanAPI project. Transforming Government: People, Process and Policy, 8 (2). pp. 205-229. ISSN 1750-6166.

Liu,Q et al., (2013). Data Provenance and Data Management Systems. In *Data Provenance and Data Management in eScience*, Springer Berlin Heidelberg.

Lopez-de-Opina, D et al., (2013). Citizen-centric Linked Data Apps for Smart Cities. Lecture *Notes in Computer Science*, Springer Publishers. pp.70-77.

Lotzmann, U. and Wimmer, M., (2012). Provenance and Traceability in Agent-based Policy Simulation. In *Proceedings of 26th European Simulation and Modelling Conference - ESM'2012*.

Luzeaux, D and Ruault, J.R., (2010). Systems of Systems. *ISTE Ltd and John Wiley & Sons Inc*.

Margo, D and Smogor, R., (2010) Using provenance to extract semantic file attributes. *In: Proceedings of the 2nd conference on Theory and practice of provenance*.

McClatchey, R et al., (2014). Provenance Support for Medical Research. In *5th International Provenance and Annotation Workshop (IPAW2014)*.

Packer, H. et al., (2014). Semantics and Provenance for Accountable Smart City Applications. *Semantic Web – Interoperability, Usability,* Applicability an IOS Press Journal.

Ram, S and Liu, J., (2009). A new perspective on Semantics of Data Provenance. First International Workshop on the role of Semantic Web in Provenance Management (SWPM).

Scherer, S. et al. (2015). Evidence Based and Conceptual Model Driven Approach for Agent-Based Policy Modelling. *Journal of Artificial Societies and Social Simulation.*

Shamdasani, J. et al. (2014). CRISTAL-ISE : Provenance Applied in Industry. Proceedings of the *16$^{th}$ International Conference on Enterprise Information Systems (ICEIS).*

Simmhan, Y.L. et al. (2005). A Survey of Data Provenance Techniques. In *Technical Report TR-618: Computer Science Department, Indiana University*.

Smarticipate, (2016). Smart services for calculated impact assessment in open governance. EC H2020 Project start February 2016.

Tsoukiàs, A. et al. (2013). Policy analytics: An agenda for research and practice. *EURO Journal on Decision Processes*, 1, 115–134.

Urban API, (2014). Interactive Analysis, Simulation and Visualisation Tools for Urban Agile Policy Implementation. Available at: http://www.urbanapi.eu/ [Accessed 5 March, 2016]

World Health Organization, (2016). Climate change and human health. Available at: http://www.who.int/globalchange/ecosystems/urbanization/en/ [Accessed 5 March, 2016]